\documentclass[preprint]{revtex4}

\usepackage[dvips]{graphicx}

\def\lp {\left( }
\def\rp {\right) }
\def\lb {\left[ }
\def\rb {\right] }
\def\lc {\left\{ }
\def\rc {\right\} }

\def\nn {\nonumber}

\def\beq{\begin{equation}}
\def\eeq{\end{equation}}
\def\bea{\begin{eqnarray}}
\def\eea{\end{eqnarray}}
\def\ni{\noindent}

\def\cL {{\cal{L}}}
\def\cM {{\cal{M}}}

\def\fp {f_\p}

\def\di {\partial_\mu }
\def\ds {\partial^\mu }

\def\cd {\!\cdot\!}
\def\sp {\!+\!}
\def\sm {\!-\!}

\def\rar {\rightarrow}

\def\a{\alpha}
\def\b{\beta}

\def\e{\epsilon}
\def\g{\gamma}

\def\l {\lambda}
\def\L {\Lambda}
\def\m{\mu}

\def\o{\omega}
\def\O{\Omega}
\def\p{\pi}

\def\s{\sigma}
\def\S{\Sigma}

\def\bpi {\mbox{\boldmath $\pi$}}
\def\bphi {\mbox{\boldmath $\phi$}}

\begin{document}

\title{chiral symmetry\\
and\\
parametrization of scalar resonances}

\author{L.O. Arantes and M.R. Robilotta}
\email{lecio@if.usp.br}
\email{robilotta@if.usp.br}
\affiliation{Instituto de F\'{\i}sica, Universidade de S\~{a}o
Paulo,\\C.P. 66318, 05315-970, S\~ao Paulo, SP, Brazil }


\date{\today}

\begin{abstract}{The linear $\s$-model is used to study the effects
of chiral symmetry in unitarized amplitudes incorporating scalar
resonances. When just a single resonance is present, we show that
the iteration of a chiral tree amplitude by means of regularized
two-pion loops preserves the smallness of $\p\p$ interaction at
low energies and estimate the importance of pion off-shell
contributions. The inclusion of a second resonance is performed by
means of a chiral extension of the linear $\s$-model lagrangian.
The new $\p\p$ ampitude at tree level complies with low-energy
theorems, depends on a mixing angle and has a zero for a given
energy between the resonance masses. The unitarization of this
amplitude by means of two-pion loops preserves both its chiral low
energy behavior and the position of this zero confirming, in a
lagrangian framework, conclusions drawn previously by T\"ornqvist.
Finally, we approximate and generalize our results and give a
friendly expression that can be used in the parametrization of $N$
coupled scalar resonances.}
\end{abstract}
\maketitle \tableofcontents

\section{introduction}

Scalar mesons have since long proved to be the most elusive states in low energy hadron physics.
At present, after decades of research, one still is not sure as how to classify them into multiplets or what their quark and gluon
contents are\cite{Cl}.
On the empirical side, one also finds important uncertainties in masses, widths, or even in the very existence of some states.

Part of the difficulties in understanding the scalar sector may be ascribed to the fact that resonances
can couple through intermediate states containing two identical pseudoscalar particles.
About ten years ago this important aspect of the problem was discussed by T\"ornqvist\cite{Tor}, who set a rather
useful and comprehensive theoretical framework for describing the role of such couplings, based on the
unitarized quark model.
The interference of resonances was also considered by Svec\cite{Sv}, using phase shifts and non-relativistic quantum
mechanics.

The interest in the scalar sector was revived recently by
evidences provided by the E791 Fermilab experiment of the
existence of resonances with low masses and large widths in the
decays $D^+ \rar (\p^- \p^+) \; \p^+$\cite{cbpfD1} and $D_s^+ \rar
(K^- \p^+) \; \p^+$\cite{cbpfD2}. The former finding was confirmed
in a number of other reactions: $D^0 \rar K_s^0 \; (\p^-
\p^+)$\cite{Cleo, Belle, BaBar1}, $\phi \rar \g\;  (\p^-
\p^+)$\cite{Kloe}, $J/\psi \rar \o\;(\p^-\p^+)$\cite{Bes}, $B^+
\rar (\p^- \p^+ )\p^+$\cite{BaBar2}. These recent results motivate
the present work, in which we discuss how chiral symmetry affects
the low-energy region of these processes and may influence  the
parameters of a light and broad resonance and its couplings to
heavier partners.

Quantum chromodynamics (QCD) is the basic theoretical framework for the study of hadronic processes,
but its non-Abelian structure hampers analytic low-energy calculations.
Therefore one needs to resort to effective theories, which mimic QCD.
In order to be really effective, these theories must be Poincar\'e invariant and possess approximate
either $SU(2)\times SU(2)$ or $SU(3)\times SU(3)$ symmetries, broken by small Goldstone boson masses.
For the sake of simplicity, we restrict ourselves to the $SU(2)$ sector.

The unitarized elastic $\p\p$ amplitudes discussed here are obtained by iterating their tree counterparts.
In section 2, we review the main features of the $\s$-model description of these building blocks and,
in section 3, derive a unitarized amplitude for the single resonance case.
As a large part of the algebraic effort needed in this result is associated with the treatment
of pion-off shell effects, in section 4 we assess their numerical importance.
In section 5 we extend the linear $\s$-model in order to allow the inclusion of a second resonance and,
in section 6, study its coupling to the first one by means of two-pion loops.
Finally, in section 7, we summarize our results and give a simple expression that can be applied in data analyses.
We have tried to make it as self contained as possible, so that it could be read directly by those people not interested
in technical details.

\section{chiral symmetry}

The intense activity on chiral perturbation theory performed in the last twenty years has made clear
the convenience of working with non-linear realizations of the symmetry.
On the other hand, when dealing with scalar resonances, one may be tempted to employ  the old
and well known linear $\s-$model.
The advantage of the former is that it is more general and incorporates all the possible freedom compatible with the symmetry.
On the other hand, it is non-renormalizable and one has to resort to order-by-order renormalization
in order to circumvent this difficulty.
The less general linear model is not affected by this problem.
As we discuss in the sequence, for a given choice of parameters, results from the linear and non-linear models become identical
at tree level.

In the framework of chiral symmetry, the inclusion of resonances
must be performed in such a way as to preserve the low-energy
theorems for $\p\p$, scattering derived by means of current
algebra. Quite generally, the amplitude $T_{\p\p}$ for the process
$ \p^a(p)\,\p^b(q) \rar \p^c(p')\,\p^d(q')$ can be written as

\beq
T_{\p\p} = \delta_{ab}\delta_{cd}\, A(s) +  \delta_{ac}\delta_{bd}\, A(t) +  \delta_{ad}\delta_{bc}\, A(u)\;,
\label{2.1}
\eeq

\ni
with $s=(p\!+\! q)^2$, $t=(p\!-\!p')^2$, $u=(p\!-\!q')^2$.
A low-energy theorem ensures that the functions $A(x)$, for $x=s,t,u$, must have the form

\beq
A(x) = \frac{x-\m^2}{f_\p^2} + \cdots  \;,
\label{2.2}
\eeq

\ni
where $\m$ and $f_\p$ are the pion mass and decay constant and the ellipsis indicates higher order contributions.
\begin{figure}[h]
    \center
    \includegraphics[scale=0.6]{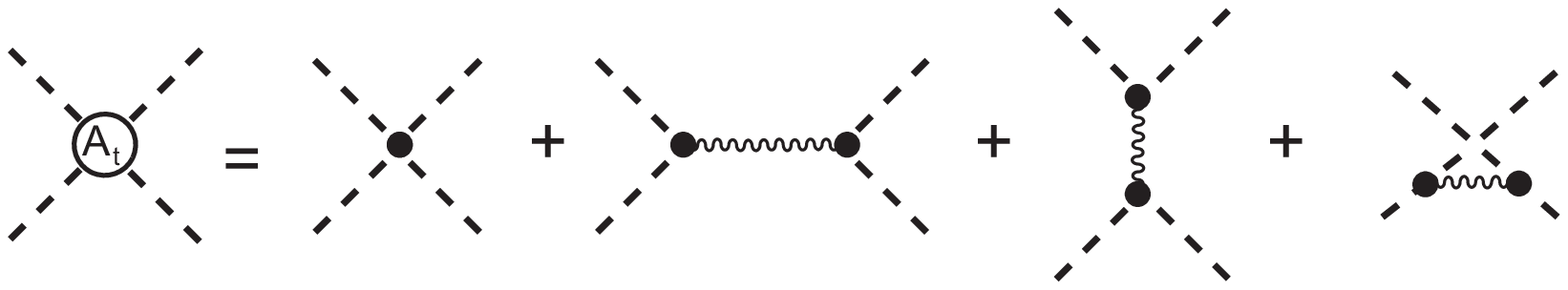}
    \caption{Tree amplitude $A_t$; dashed and thin wavy lines
represent pions and a scalar resonance.}
\end{figure}
\ni


When a scalar-isoscalar resonance is present, the tree level
amplitude for $\p\p$ scattering is given by the four diagrams of
fig.1, irrespectively of whether the symmetry is implemented
linearly or not. We begin by considering the linear $\s-$model,
described by the lagrangian

\beq
\cL_\s =  \frac{1}{2}\lp \di \s \; \ds \s  + \di \bpi \cd \ds \bpi \rp
-\frac{m^2}{2} \lp \s^2 \!+\! \bpi^2 \rp - \frac{\l}{4} \lp \s^2 \!+\! \bpi^2 \rp ^2 + c \,  \s \;.
\label{2.3}
\eeq

Denoting by $f$ the fluctuations of the scalar field and using $\s = f_\p \!+\! f$, one finds, at tree level,

\beq
\m^2 = m^2 + \l f_\p^2 \;, \;\;\;\;\;\; c=  \m^2 f_\p \;, \;\;\;\;\;\; M_\s^2 = 2 \l f_\p^2 + \m^2 \;,
\label{2.4}
\eeq

\ni
$M_\s$ being the $\s$ mass.
The $\p\p$ scattering amplitude is

\beq
A_t(x) = -2\;\l -\; \frac{4\;\l^2\; f_\p^2}{x-M_\s^2}\;,
\label{2.5}
\eeq

\ni
where the subscript $t$ stands for {\em tree} and
the two contributions on the r.h.s. arise respectively from the four-pion vertex and one of the resonance terms in fig.1.
Comparing this result with eq.(\ref{2.2}), one learns that none of these contributions is isolatedly compatible with
the low-energy theorem.
However, when both terms are added, one has

\beq
A_t(x) = \frac{x\! -\! \m^2}{f_\p^2} \; \lb 1- \;\frac{x \sm \m^2}{x \sm M_\s^2}\rb\;,
\label{2.6}
\eeq

\ni and consistency becomes explicit, since $M_\s^2 >> \m^2 \sim x$.
This result conveys an important message, namely that, in the linear
model, the resonance and the non-resonating background must always
be treated in  the same footing, for the sake of  preserving chiral
symmetry. As we discuss in the sequence, this issue is especially
relevant for the definition of the resonance width.

In the alternative approach, the scalar field $f$ couples to pion fields $\bphi$,
which behave non-linearly under chiral transformations\cite{Wei}.
In this new framework, the field $f$ is assumed to be a true chiral scalar, invariant under both vector and axial transformations,
and should not be confused with $\s$, the chiral partner of the pion in the linear $\s$-model.
The effective lagrangian for this system is written as\cite{MR}

\bea
\cL &=& \frac{1}{2}\lp \di f  \; \ds f  - M_\s^2 f^2 \rp
+ \frac{1}{2f_\p} \lp f_\p + c_\chi f  \rp \lp \di \bphi \cd \ds \bphi
+\di \sqrt{\fp^2-\bphi^2}\; \ds \sqrt{\fp^2-\bphi^2} \rp
\nn\\[2mm]
&+& \m^2 \lp f_\p + c_bf  \rp \sqrt{\fp^2-\bphi^2}\;,
\label{2.7}
\eea

\ni
where the dimensionless constants $c_\chi$ and $c_b$ represent, respectively, the scalar-pion couplings
that preserve and break chiral symmetry.

The evaluation of the diagrams of fig.1 then yields

\beq
A_t(x) =  \frac{x-\m^2}{f_\p^2} - \frac{c_\chi^2/4 \; [ (x-\m^2) + \e\; \m^2 ]^2}{f_\p^2\;(x-M_\s^2)}\;
\label{2.8}
\eeq

\ni
where $\e=2c_b/c_\chi-1$ and,  as before, the two contributions are due respectively to the four-pion vertex and to the resonance.
In this case, however, each of the contributions conforms independently with the low-energy theorems.
The former gives rise to the leading term of eq.(\ref{2.2}) and the latter corresponds to a higher order correction.
This result sheds light into the role of a resonance in the framework of chiral symmetry.
We note that, for $c_\chi=2$ and $\e=0$, one recovers the result from the linear $\s-$model, given by eq.(\ref{2.6}).
The non-linear lagrangian gives rise to more general results, since they hold for any choices
of the parameters $c_\chi$ and $c_b$.
On the other hand, it is not renormalizable, because the coupling constant $c_\chi/2f_\p$ carries a negative dimension.

With future purposes in mind, we rewrite the result from the linear model as

\beq
A_t(x) = - \; \frac{\g^2}{x-M_\s^2}\;,
\label{2.9}
\eeq

\ni
with

\beq
\g^2(x) =  (x\!-\!\m^2)(M_\s^2\!-\!\m^2) / f_\p^2\;.
\label{2.10}
\eeq

In the evaluation of the effects of pion loops, it is useful to associate diagrams directly with eq.(\ref{2.9}).
We do this by reexpressing the $\p\p$ amplitude of fig.1 as in fig.2, where the thick wavy lines now include
the contribution from the four-pion contact interaction and the function $\g(x)$ implements the effective couplings at the vertices.
\begin{figure}[h]
    \center
    \includegraphics[scale=0.6]{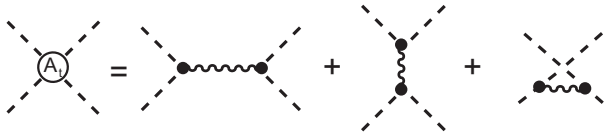}
    \caption{Tree amplitude $A_t$; the thick wavy lines
incorporate the contact term of fig.1.}
\end{figure}


\section{s-channel loops}

We work in the linear model and construct the dynamical features of the scalar resonance
by considering only iterated contributions from a single loop.
In this approximation, the dressed propagator is determined by the three diagrams shown in fig.3a.
The last of them corresponds to a composite Dyson series and includes all possible iterations of the $\p\p$
tree amplitude, as represented in fig.3b.
\ni
\begin{figure}[h]
    \center
    \includegraphics[scale=0.6]{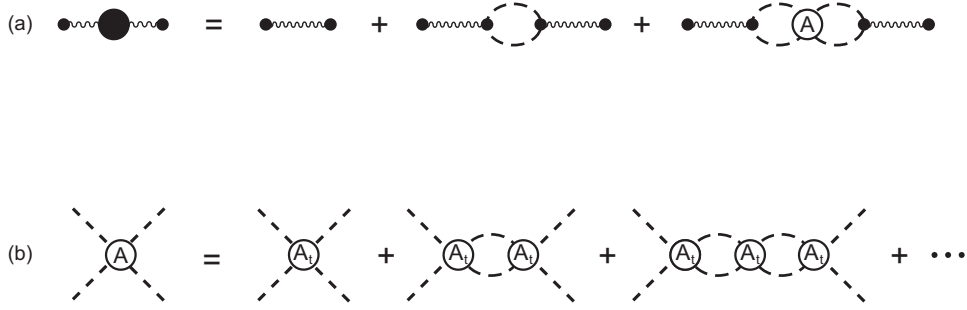}
    \caption{(a) Full resonance propagator;  (b) $s$-channel
unitarized $\p\p$ amplitude.}
\end{figure}


In this work we are mostly interested in exploring the behavior of
coupled resonances. With this purpose in mind, we make a
simplifying approximation and consider only the amplitude
associated with the first diagram on the r.h.s. of fig.2, which is
denoted by $A_t \equiv A_t(s)$ and given by eq.(\ref{2.9}), for
$x=s$. It is worth recalling, however, that the diagrams in the
$t$ and $u$ channels also do play a visible role, as discussed in
refs.\cite{BBWL} and \cite{AS}. The single loop contribution to
the $\p\p$ scattering amplitude is given by

\beq
A_1(s) = A_t \lb -\O \rb A_t \;,
\label{3.1}
\eeq

\ni
where the function

\beq
\O(s) = -\; \frac{3}{32\p^2} \lb L + \L_\infty \rb
\label{3.2}
\eeq

\ni
contains an infinite constant $\L_\infty$ and a finite component $L(s)$.
The latter can be evaluated analytically and is given by

\bea
&& \bullet \; 0\leq s < 4 \m^2 \rar L(s)
= -\;2\; \frac{\sqrt{4 \m^2-s}}{\sqrt{s}}\; \tan^{-1} \lb \frac{\sqrt{s}}{\sqrt{4 \m^2-s}}\rb \;,
\label{3.3}\\[2mm]
&& \bullet \; 4 \m^2 \leq s \rar L(s)
=  \frac{\sqrt{s - 4\m^2}}{\sqrt{s}}\;\lc
\ln \lb \frac{\sqrt{s} - \sqrt{s-4\m^2}}{\sqrt{s} + \sqrt{s-4\m^2}}\rb  + i\;\p \rc \;.
\label{3.4}
\eea

The behavior of the function $L(s)$ is displayed in fig. 4, where
it is possible to notice a cusp at $s= 4\m^2$.
\begin{figure}[h]
    \center
    \includegraphics[scale=0.4]{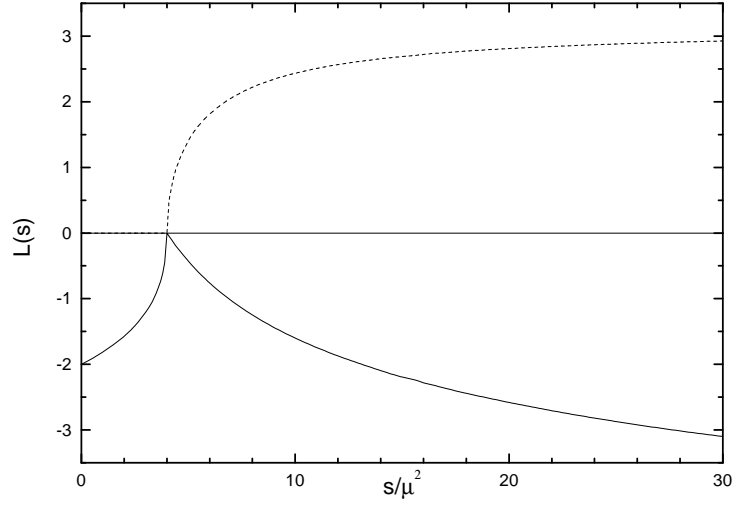}
    \caption{Function $L(s)$, that determines the self energy associated with the loop.}
\end{figure}

In the linear $\s$-model beyond tree level, loops bring infinities
which must be removed consistently. The renormalization of the
$\s$-model was discussed by Lee and collaborators \cite{BBWL, BWL}
and reviewed in a pedagogical way in ref.\cite{PS}. In order to keep
only the essential features of our discussion, we note that the
dynamical scalar mass can be cut along a $\p\p$ loop, whereas the
pion mass can be cut along a $\p\s$ loop. As the latter is heavier,
we assume that changes in the pion mass can be neglected at the
energy scale one is working at. The lifting of  this restriction is
straightforward, but would require a considerable increase in the
algebraic effort. Since at one-loop level the wave function
renormalization is finite\cite{PS}, the elimination of $\L_\infty$
from eq.(\ref{3.2})  is performed by making $m\rar m_0$ and $\l \rar
\l_0$ in the linear lagrangian and rewriting it as

\bea
\cL &=&  \frac{1}{2}\lp \di \s \; \ds \s  + \di \bpi \cd \ds \bpi \rp
-\frac{m^2}{2} \lp \s^2 \!+\! \bpi^2 \rp - \frac{\l}{4} \lp \s^2 \!+\! \bpi^2 \rp ^2 + f_\p\,\m^2 \, \s
\nn\\[2mm]
&-& \frac{\delta_m}{2} \lp \s^2 \!+\! \bpi^2 \rp - \frac{\delta_\l}{4}   \lp \s^2 \!+\! \bpi^2 \rp^2 \;,
\label{3.5}
\eea

\ni
with $\delta_m = m_0^2 - m^2$ and $\delta_\l = \l_0 - \l$.
E expanding $\s$ around $f_\p$,
using the condition $\delta_m\! =\!-f_\p^2\, \delta_\l$  associated with the constancy of $\m^2$
and noting that tadpoles do not contribute by construction\cite{PS}, we find

\bea
\cL &=& \frac{1}{2}\lp \di f \; \ds f - M_\s^2 f^2 \rp
+ \frac{1}{2} \lp \di \bpi \cd \ds \bpi - \m^2 \bpi^2 \rp
- \l \, f_\p \,f\,\bpi^2 - \l\, \bpi^4 +\cdots
\nn\\[2mm]
&-& \delta_\l \lp f_\p^2 \, f^2 + f_\p \, f \, \bpi^2 + \bpi^4/4 + \cdots \rp \;.
\label{3.6}
\eea

This result gives rise to the counterterm diagrams shown in fig.
5, which allow the factor $\L_\infty$ in eq.(\ref{3.2}) to be
killed by a suitable choice of $\delta_\l$. We are then entitled
to replace $\O(s)$ in eq.(\ref{3.1}) by

\beq
\bar{\O}(s) = -\; \frac{3}{32\p^2} \lb L + c \rb \;,
\label{3.7}
\eeq

\ni where $c$ is a yet undetermined constant. Denoting by
$\bar{R}$ and $I$ the real and imaginary parts of $\bar{\O}$, the
usual self energy insertion is written as

\beq
\bar{\Sigma}(s) = \g^2 \lb \bar{R} + i\;I \rb \;.
\label{3.8}
\eeq
\ni
\begin{figure}[h]
    \center
    \includegraphics[scale=0.6]{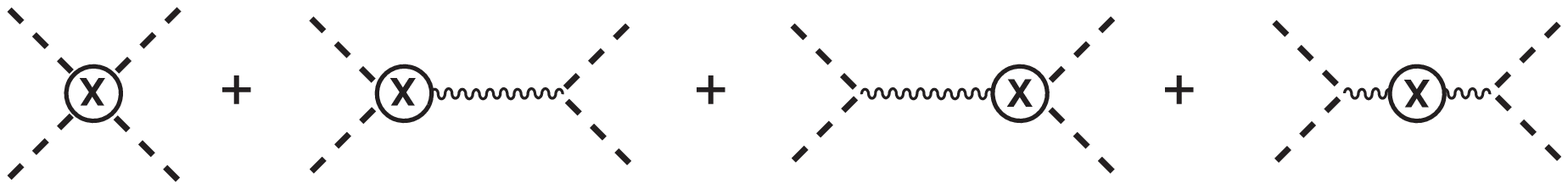}
    \caption{Counterterm structure for $A_1(s)$.}
\end{figure}

Considering all possible iterations of the two-pion loop, we construct the full $s-$channel $\p\p$ amplitude given in fig. 3b.
This geometrical series can be summed and one finds

\beq
\bar{A}(s) = -\; \frac{\g^2} {s-\cM_A^2 + i\;M_\s \;\Gamma_A} \;,
\label{3.9}
\eeq

\ni
with $\cM_A^2(s) = M_\s^2 + \g^2 \bar{R}$  and $M_\s \Gamma_A(s) = \g^2 I$.
The scalar propagator, fig.3a, can be regularized by the same set of couterterms and reads

\beq
\bar{\Delta}(s) = \frac{1} {s-\cM_\Delta^2 + i\;M_\s \;\Gamma_\Delta} \;,
\label{3.10}
\eeq

\ni
where

\bea
&& \cM_\Delta^2(s) = \m^2
+ \frac{f_\p^2 (M_\s^2 \sm \m^2) [f_\p^2 \sm (M_\s^2 \sm \m^2) \bar{R}]}
{ [f_\p^2 \sm (M_\s^2 \sm \m^2) \bar{R}] ^2 + (M_\s^2 \sm \m^2)^2 I^2}\;,
\label{3.11}\nn\\[2mm]
&& M_\s \Gamma_\Delta(s) =  -\; \frac{ f_\p^2 (M_\s^2 \sm \m^2)^2 \;I}
{ [f_\p^2 \sm (M_\s^2 \sm \m^2) \bar{R}] ^2 + (M_\s^2 \sm \m^2)^2 I^2}\;.
\label{3.12}\nn\\[2mm]
\eea

The amplitude $\bar{A}$ and the propagator $\bar{\Delta}$ thus yield inequivalent definitions for the resonance mass and width,
which correspond to different prescriptions for the determination of the parameter $c$ in eq.(\ref{3.7}).
We fix this constant by using the result for the $\p\p$ amplitude, for it is closer to observation.
Imposing that the pole of $\bar{A}$ occurs at the physical mass $M_\s$, one finds
$\bar{R}(M_\s^2) =0 \rar c= - \Re L(M_\s^2)$ and
the running mass becomes $\cM_A^2(s) = M_\s^2 + \g^2 [\bar{R}(s) - \bar{R}(M_\s^2)]$,
whereas the width reads

\beq
\Gamma_A(s) = \frac{3 (s \sm \m^2) (M_\s^2\!-\!\m^2)}{32\p \;f_\p^2}
\; \frac{\sqrt{s-4\m^2}}{M_\s \sqrt{s}}\; \Theta(s\!-\!4\m^2) \;.
\label{3.13}
\eeq

The signature of chiral symmetry in this problem is the factor $(s \sm \m^2)/f_\p^2$, present in the functions $\g^2(s)$
and $\bar{\S}(s)$.
It implements the low energy theorem and is due to the use of eq.(\ref{2.6}) as the main building block in the calculation.
If one were to keep just the second term of eq.(\ref{2.5}) in the evaluation of the two-pion loop contribution,
it would be replaced by $(M_\s^2 \sm \m^2)/f_\p^2$.
Thus, both procedures yield identical results at the pole, but correspond to rather different forms for the resonance width.
\ni
\begin{figure}[h]
    \center
    \includegraphics[scale=0.6]{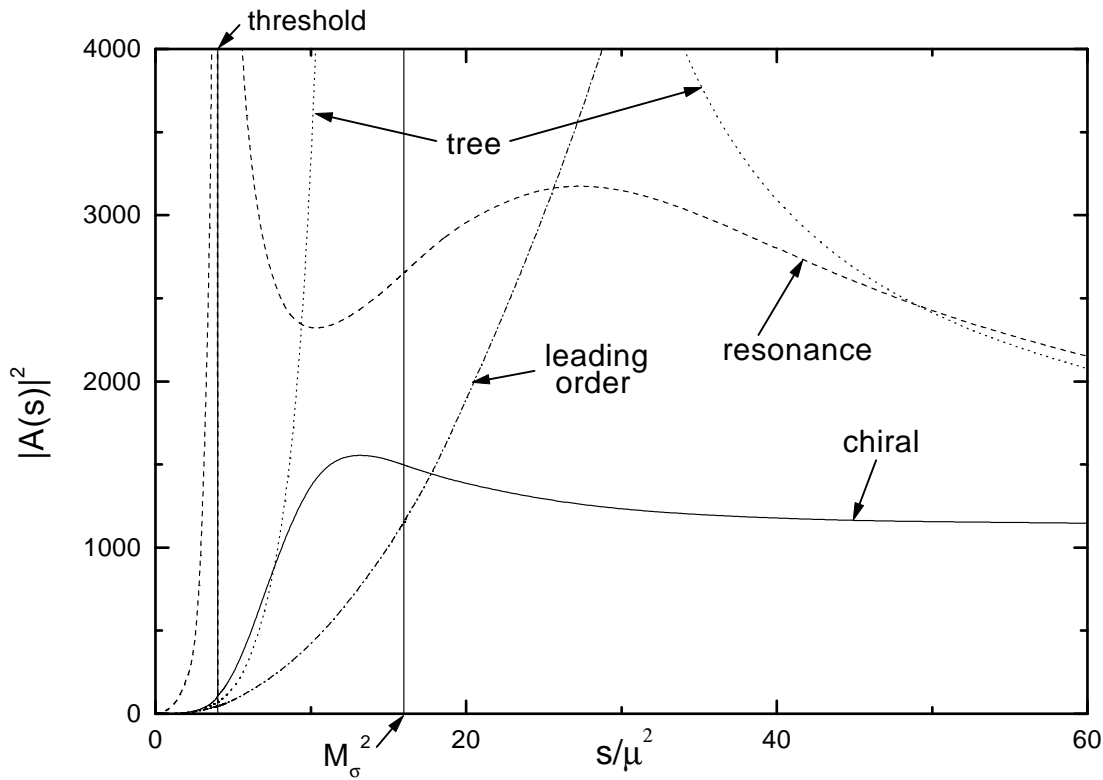}
    \caption{The functions $A(s)$ are the $\p\p$
amplitudes given by equations (\ref{2.2}) (dot-dashed line),
(\ref{2.6}) (dotted line), (\ref{3.9}) (continuous line) and by
unitarizing just the $\s$ (dashed line).}
\end{figure}

In fig.6 we explore this this aspect of the problem, in the case of the function $| A(s)|^2$, for the choice $M_\s= 4\m$.
The use of eq.(\ref{2.2}) yields the {\em leading order} curve, an unbound parabola which blows up at large energies.
The inclusion of the resonance as in eq.(\ref{2.6}) gives rise to the {\em tree} curve.
The {\em chiral} curve, given by eq.(\ref{3.9}), is obtained by iterating the {\em tree} amplitude by means of
two-pion loops.
Finally, the {\em resonance} curve is derived by iterating just the second term of eq.(\ref{2.5}) and then adding the first one.
Inspecting this figure, one learns that the last procedure violates badly chiral symmetry, since it gives rise to a result
which does not tend to the {\em leading order} one when $s\rar 0$, as predicted by the low-energy theorems.

The reason for this kind of deviation can be found in fig.7, which
shows the behaviors of the real and imaginary parts of the {\em
chiral} and {\em resonance} amplitudes, together with the
corresponding {\em leading order} and {\em tree} contributions. It
is possible to notice that, at low-energies, the {\em leading
order, tree} and {\em chiral} results stay close together,
indicating that loop contributions are small. On the other hand,
when one iterates just the second term of eq.(\ref{2.5}), loop
contributions are rather large and compatibility with the
low-energy theorem is lost.
\ni
\begin{figure}[h]
    \center
    \includegraphics[scale=0.6]{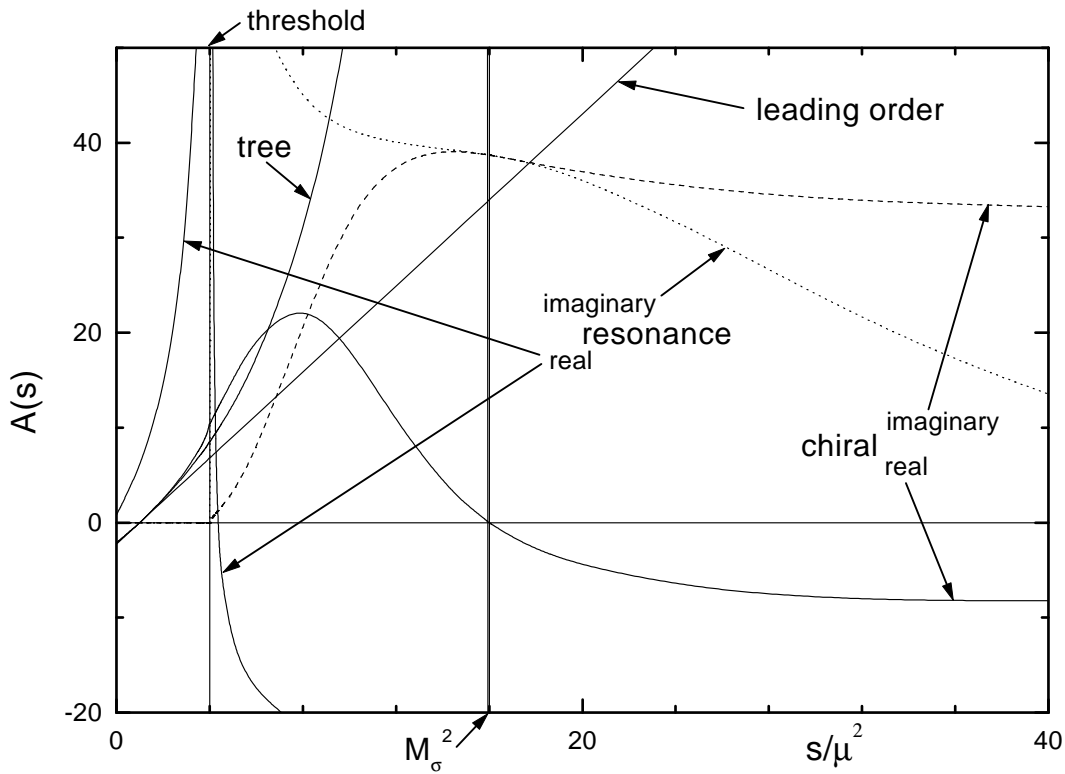}
    \caption{Real and imaginary parts of  the $\p\p$ amplitude
$A(s)$; the meanings of the labels are the same of fig.6.}
\end{figure}
\clearpage

\section{$K$-matrix unitarization}

A popular alternative procedure for unitarizing amplitudes  is based on the so called $K$-matrix formalism.
A resonance has a well defined isospin and it is useful to rewrite the generic $\p\p$ scattering amplitude as

\beq
T_{\p\p} = T_0(s,t,u)\;P_0 +  T_1(s,t,u)\;P_1 +  T_2(s,t,u)\;P_2 \;,
\label{4.1}
\eeq

\ni
where $P_I$ is the projector into the channel with total isospin $I$.
The amplitudes $T_I$ are translated into the $A(x)$ of eq.(\ref{2.1})  by\cite{GL}

\beq
T_0 = 3\; A(s) + A(t) + A(u)\;,
\;\;\;\;\;\;
T_1 = A(t) - A(u)\;,
\;\;\;\;\;\;
T_2 = A(t) + A(u)\;.
\label{4.2}
\eeq

In this work we neglect $t$ and $u$ channel effects and the scalar-isoscalar non-relativistic kernel for identical
particles is related to the relativistic tree amplitude by

\beq
K(s) = \frac{3}{2} \;\frac{A_t}{8\p\sqrt{s}}\;.
\label{4.3}
\eeq

The on-shell iteration of this kernel yields the scattering amplitude $f$, which is given by

\beq
f = K / (1-iqK) \;,
\label{4.4}
\eeq

\ni
where $q = \sqrt{s/4-\m^2}$ is the center of mass momentum.
Using $qK=\tan\delta$, one finds the usual phase shift parametrization for $f$.
The relativistic counterpart of (\ref{4.4})  reads

\beq
\bar{A}_K(s) = \frac{A_t}{1 - i (3\sqrt{s \sm 4\m^2}\;A_t / 32\p\sqrt{s})}\;.
\label{4.5}
\eeq

\ni
and, using eq.(\ref{2.9}), we find

\beq
\bar{A}_K(s) = -\;\frac{\g^2}{s-M_\s^2 +  i\;M_\s \;\Gamma_A} \;.
\label{4.6}
\eeq

In other words, one recovers the amplitude $\bar{A}(s)$ given by (\ref{3.9}), with $\bar{R}=0$.
This is expected since, as it is well known, $K\!\!-$matrix unitarization gives rise to a width,
but does not renormalize the mass.
In fig.8 we compare the functions $|\bar{A}(s)|^2$ and $|\bar{A}_K(s)|^2$, in order to show that the $K\!\!-$matrix
formalism does produce a rather decent approximation for the explicit loop calculation, at a considerably lower
algebraic cost.

\ni
\begin{figure}[h]
    \center
    \includegraphics[scale=0.6]{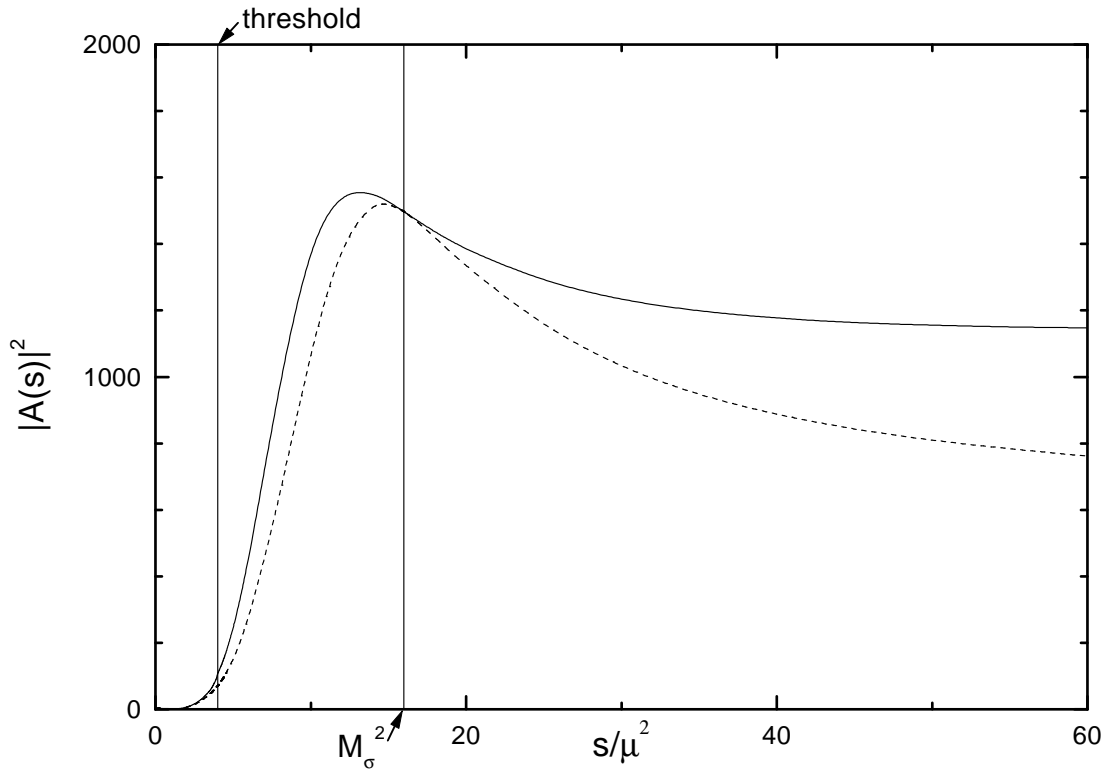}
    \caption{ The functions $A(s)$ are the $\p\p$ amplitudes given by equations (\ref{3.9}) (continuous line)
and (\ref{4.6}) (dashed line).}
\end{figure}


\section{extended $\s-$model}

We now consider the problem of generalizing the linear $\s-$model, so that it could encompass two resonances.
With this purpose in mind, we introduce a second scalar-isoscalar field $\xi$, which is assumed to be a chiral scalar.
In other words, this new field is invariant under both isospin and axial transformations of the group $SU(2)\times SU(2)$.
This allows its physical content to be compatible with realizations outside the $SU(2)$ sector such as,
for instance, $s\bar{s}$ or glueball states.

In order to preserve renormalizability, we avoid couplings with
negative dimensions and add two new chiral invariant terms to the
$\cL_\s$ of eq.(\ref{2.3}). The two-resonance lagrangian becomes

\beq
\cL_{\s\xi} = \cL_\s + \frac{1}{2}\lp \di \xi \; \ds \xi - M_\xi^2 \, \xi^2 \rp
+ g \,\xi\, \lp \s^2+\bpi^2\rp \;,
\label{5.1}
\eeq

\ni
where $M_\xi$ is the $\xi$ mass and $g$ is a coupling constant.
When the $\s$ is reexpressed in terms of the fluctuation $f$, the new interaction lagrangian gives rise
to a contribution linear in $\xi$, indicating that this field also has a classical component, denoted by $e$.
Writing $\s=f_\p + f$ and $\xi = e +\e$, we find

\bea
\cL_{\s\xi} &=& \lb -(m^2/2 \sm ge) f_\p^2 \sm \l f_\p^4/4 \sp c f_\p \rb
+ \lb - (m^2 \sm 2ge) f_\p \sm \l f_\p^3 \sp c\rb f
\nn\\[2mm]
&+& \frac{1}{2}\lb \di \bpi \cd \ds \bpi  - (m^2 \sm 2ge \sp \l f_\p^2) \,\bpi^2 \rb
+ \frac{1}{2}\lb \di f \; \ds f - (m^2 \sm 2ge \sp 3\l f_\p^2) \,f^2 \rb
\nn\\[2mm]
&-& \lb \l f_\p f (f^2+\bpi^2) + \l \; \bpi^4/4 + \cdots \rb
+ \lb -M_\xi^2 e \sp g f_\p^2 \rb \e
\nn\\[2mm]
&+& \frac{1}{2}\lp \di \e \; \ds \e - M_\xi^2 \, \e^2 \rp
+ g \,\e\, \lp f^2+\bpi^2\rp + 2 \, g\, f_\p\, f\, \e \;.
\label{5.2}
\eea

The conditions $[- (m^2 \sm 2ge) f_\p \sm \l f_\p^3 \sp c] = 0$ and $[-M_\xi^2 e \sp g f_\p^2] =0$
for the free parameters allow the elimination of the linear terms in $f$ and $\e$.
The $\bpi$ and $\s$ masses are

\beq
\m^2 = m^2 \sm 2 g e \sp \l f_\p^2 \;,
\;\;\;\;\;\;
M_\s^2 = \m^2 \sp 2 \l f_\p^2\;.
\label{5.3}
\eeq

The last term in eq.(\ref{5.2}) corresponds to a mass mixing, which is eliminated by introducing new fields $\a$ and $\b$,
given by

\beq
\a = \cos\theta \,f  + \sin\theta \, \e \;,
\;\;\;\;\;\;
\b = - \sin \theta \, f + \cos\theta \,\e \;,
\label{5.4}
\eeq

\ni
and choosing the angle $\theta$ such that $\tan 2\theta = 4 g f_\p /(M_\xi^2 \sm M_\s^2)$.
This yields

\beq
\cos^2\theta \; M_\a^2 + \sin^2\theta \; M_\b^2 = M_\s^2 \;,
\;\;\;\;\;\;
\sin^2\theta \; M_\a^2 + \cos^2\theta \; M_\b^2 = M_\xi^2 \;.
\label{5.5}
\eeq

\ni
and allows the lagrangian to be written as

\bea
\cL_{\s\xi} &=& \frac{1}{2}\lb \di \bpi \cd \ds \bpi  - \m^2 \,\bpi^2 \rb
+ \frac{1}{2}\lb \di \a \; \ds \a - M_\a^2 \, \a^2 \rb
+ \frac{1}{2}\lb \di \b \; \ds \b - M_\b^2 \, \b^2 \rb
\nn\\[2mm]
&-& f_\p \lp \l_\a \;\a + \l_\b \; \b \rp \bpi^2 - \l\;\bpi^4 /4 + \cdots
\label{5.6}
\eea

\ni
where the coupling constants $\l_\a$, $\l_\b$ and $\l$ are completely determined by the masses and mixing angle as

\bea
&& \l_\a = \cos\theta \;\lp M_\a^2 \sm \m^2 \rp / 2 f_\p^2 \;,
\;\;\;\;\;\;
\l_\b = -\sin\theta \;\lp M_\b^2 \sm \m^2 \rp / 2 f_\p^2 \;,
\label{5.7}\\[2mm]
&& \l = \lb \cos^2\theta \lp M_\a^2 \sm \m^2 \rp + \sin^2\theta \lp M_\b^2 \sm \m^2 \rp \rb / 2 f_\p^2 \;.
\label{5.8}
\eea

The tree amplitude for $\p\p$ scattering is given by the diagrams of fig.9 and reads

\bea
&& A_t(x) = -2\;\l -\; \frac{4\;\l_\a^2\; f_\p^2}{x-M_\a^2}
-\; \frac{4\;\l_\b^2\; f_\p^2}{x-M_\b^2}
\nn\\[2mm]
&&
=  \frac{x\! -\! \m^2}{f_\p^2} \;
\lb 1- \cos^2\theta \; \frac{x \sm \m^2}{x-M_\a^2} - \sin^2\theta \; \frac{x \sm \m^2}{x-M_\b^2}\rb\;.
\label{5.9}
\eea

\ni
\begin{figure}[h]
    \center
    \includegraphics[scale=0.6]{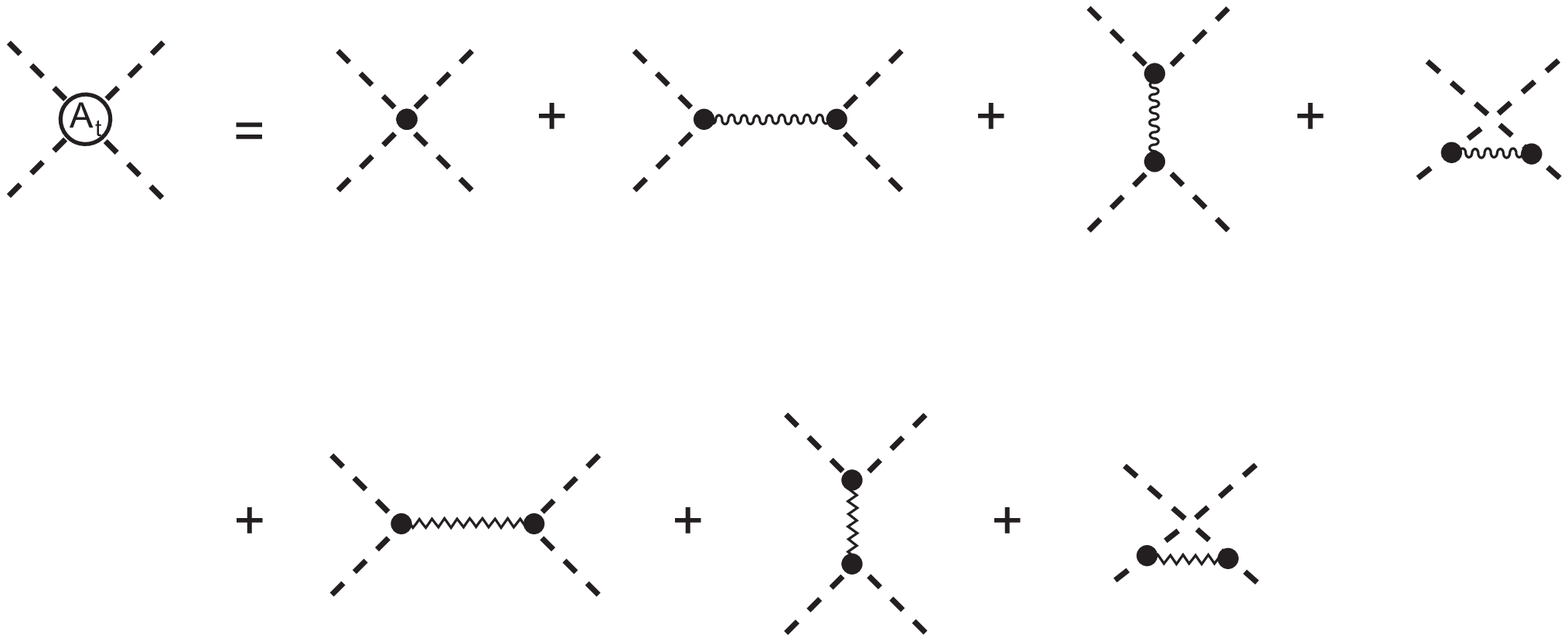}
    \caption{Tree amplitude $A_t$; dashed and thin wavy and zigzag lines represent pions and scalar resonances
$\a$ and $\b$.}
\end{figure}


This result corresponds to the generalization of eq.(\ref{2.6}) and is consistent, as it must be, with the low energy theorem.
As in the single resonance case, it is convenient to write the tree amplitude as

\beq
A_t(x) \equiv A_{t\a}(x) + A_{t\b}(x) =
-\; \frac{\g_\a^2}{x-M_\a^2}
- \; \frac{\g_\b^2}{x-M_\b^2}\;,
\label{5.10}
\eeq

\ni
with

\beq
\g_\a^2(x) = \cos^2\theta \, (x \sm \m^2) (M_\a^2 \sm \m^2) / f_\p^2 \;,
\;\;\;\;\;\;\
\g_\b^2(x) = \sin^2\theta \, (x \sm \m^2) (M_\b^2 \sm \m^2) / f_\p^2 \;,
\label{5.11}
\eeq

\ni
and reexpress the diagrams of fig.9 as in fig.10, where the thick lines now incorporate the contributions from
the four-pion contact interaction and the functions $\g_i^2$ correspond to effective couplings.

.
\ni
\begin{figure}[h]
    \center
    \includegraphics[scale=0.6]{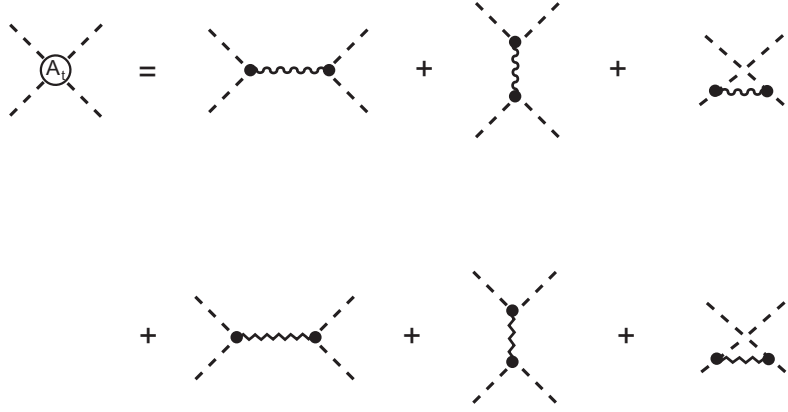}
    \caption{Tree amplitude $A_t$; the thick wavy and zigzag lines
    incorporate the contact term of fig.9.}
\end{figure}

\section{coupled resonances}

In the case of two scalar resonances $\a$ and $\b$, which can
couple through a two-pion intermediate state, one has to consider
the four two-point functions displayed in fig.11a. The structures
of these functions are given in figs.11b and depend on the full
elastic $\p\p$ amplitude.

\ni
\begin{figure}[h]
    \center
    \includegraphics[scale=0.6]{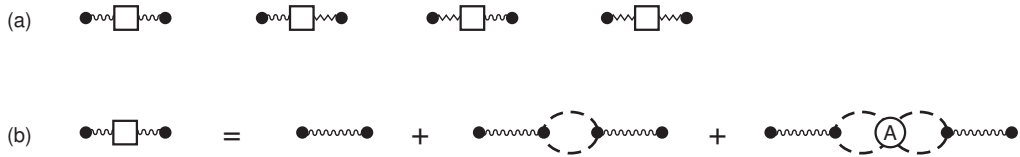}
    \caption{(a) Coupled  resonance propagators and (b) their
dynamical structures; dashed and thin wavy and zigzag lines
represent pions and scalar resonances $\a$ and $\b$.}
\end{figure}

As in the single resonance case, the $\p\p$ amplitude is obtained by iterating the tree result from the previous section.
The first iteration of eq.(\ref{5.10}) yields

\beq
A_1(s) = \lb A_{t\a} + A_{t\b} \rb \lb - \O \rb  \lb A_{t\a} + A_{t\b} \rb \;,
\label{6.1}
\eeq

\ni where $\O$ is given by eq.(\ref{3.2}) and contains a
divergence that needs to be removed by renormalization. The same
formal manipulations used in section 3 allow couterterms to be
generated in the two-resonance lagrangian, eq.(\ref{5.1}),  and
the regularized version of $A_1$ reads

\beq
\bar{A}_1(s) = \sum_{i=\a}^\b  \sum_{j=\a}^\b  A_{ti} \lb - \bar{\O}_{ij} \rb  A_{tj} \;,
\label{6.2}
\eeq

\ni
with

\beq
\bar{\O}_{ij}(s) = -\; \frac{3}{32\p^2} \lb L + c_{ij}  \rb \;.
\label{6.3}
\eeq

The self-energy associated with a particular interaction is given by

\beq
\bar{\Sigma}_{ij}(s)  = \g_i  \;\g_j  \lb \bar{R}_{ij}+ i\;I \rb  \;.
\label{6.4}
\eeq

\ni
\begin{figure}[h]
    \center
    \includegraphics[scale=0.6]{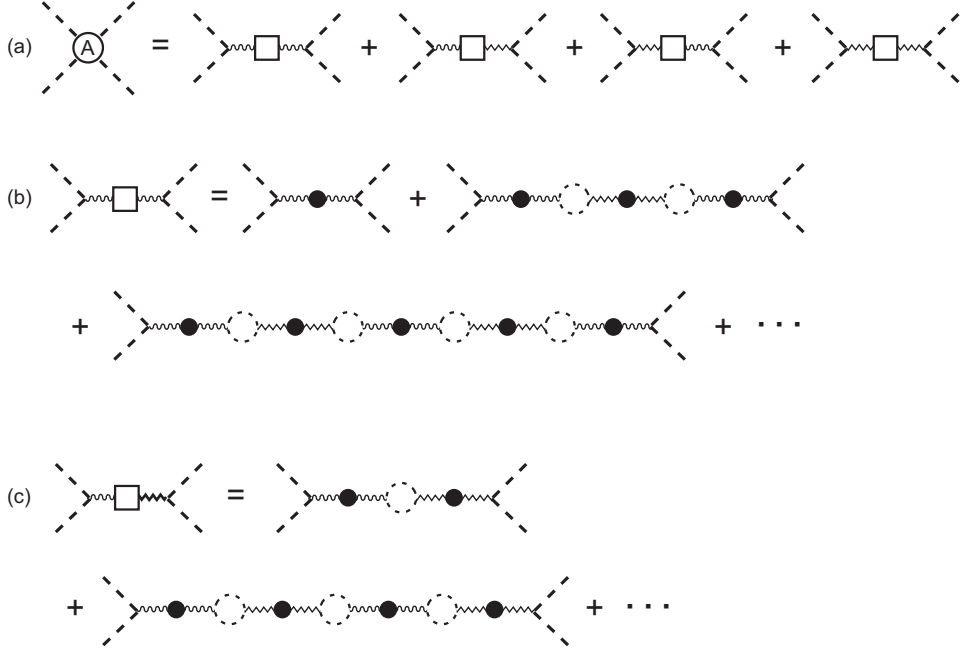}
    \caption{(a) Coupled  resonance contribution to the $\p\p$ amplitude and (b, c) partial contributions.
The meaning of thick wavy and zigzag lines is given in fig. 10.}
\end{figure}

The iteration of this amplitude to all orders gives rise to the structure shown in fig.12a, which contains four sub amplitudes,
denoted by $\bar{A}_{ij}$.
In order to construct these functions, we first evaluate the single resonance contributions from fig.12b,
and recover result given in eq.(\ref{3.9}).
We then assemble all possible combinations of these results, as in figs.12c, and find the diagonal and off-diagonal
amplitudes as

\bea
&& \bar{A}_{\a\a}(s) = \frac{-\g_\a^2\; [s \sm M_\b^2 \sm \g_\b^2 (\bar{R}_{\b\b}+ i I)]}
{D + i\;G}\;.
\label{6.5}\\[2mm]
&&\bar{A}_{\a\b}(s) = \frac{-\g_\a^2 \; \g_\b^2 \; ( \bar{R}_{\a\b}+ iI)}
{D + i \;G }\;.
\label{6.6}
\eea
\ni
with

\bea
D(s) &=& \lp s\!-\!M_\a^2\!-\!\g_\a^2 \, \bar{R}_{\a\a}\rp  \lp s\!-\!M_\b^2\!-\!\g_\b^2 \, \bar{R}_{\b\b}\rp
- \g_\a^2\g_\b^2 \lp \bar{R}_{\a\b}\rp ^2 \;,
\label{6.7}\\[2mm]
G(s) &=& - \g_\a^2\, (s\!-\!M_\b^2) - \g_\b^2\, (s\!-\!M_\a^2)
+ \g_\a^2 \, \g_b^2 \, (\bar{R}_{\a\a}+ \bar{R}_{\b\b} - 2 \, \bar{R}_{\a\b}) \;.
\label{6.8}
\eea

The expression for $\bar{A}_{\b\b}$ is obtained by making $(\a \leftrightarrow \b)$ in eq.(\ref{6.5}).
The evaluation of the full $s-$channel $\p\p$ amplitude produces

\beq
\bar{A}(s) = \frac{ G}{D + i \;G\;I}\;.
\label{6.9}
\eeq

This result allows the construction of resonance propagators.
However, the resulting expressions are rather messy and will not be quoted.
In order to determine the couterterms $c_{ij}$ in eq.(\ref{6.3}), we use directly the $\p\p$ amplitude.
Imposing that the resonances decouple at their poles, we find
$(\bar{R}_{\a\a}+ \bar{R}_{\b\b} - 2 \, \bar{R}_{\a\b})=0$.
The function $G(s)$ becomes proportional to the tree amplitude $A_t(s)$ given by eq.(\ref{5.10})  and
the unitarized amplitude can be written as

\beq
\bar{A}(s) = \frac{ A_t(s)}{[D/(s \sm M_\a^2) (s \sm M_\b^2)] + i \; A_t(s) \;I}\;.
\label{6.10}
\eeq

This result shows that the zeroes of $ \bar{A}(s)$ and $A_t(s)$ coincide,
enforcing the {\em theorem} given by T\"ornqvist\cite{Tor}, which states that
{\em "a zero in the partial wave amplitude in the physical region remains a zero after unitarization"}.
The zeroes of $A_t(s)$ occur at $s=\m^2$ and the point

\beq
s_1 = \frac{M_\a^2\,M_\b^2 - \m^2 (M_\b^2 \cos^2 \theta \sp M_\a^2 \sin^2 \theta )}
{M_\a^2 \cos^2 \theta + M_\b^2 \sin^2 \theta -\m^2}\;,
\label{6.11}
\eeq

\ni
with $M_\a^2 < s_1 < M_\b^2$.
In principle, the position of this point could be obtained from analyses of empirical data and the value of the mixing
angle $\theta$ would be related to the masses by

\beq
\tan^2 \theta = \frac{(M_\b^2 \sm s_1)\,(M_\a^2 \sm \m^2)}{(s_1 \sm M_\a^2)\, (M_\b^2 \sm \m^2)}\;.
\label{6.12}
\eeq

Imposing $D(M_\a^2) = D(M_\b^2)=0$, one finds the conditions

\bea
&& c_{\a\a}\sm c_{\b\b} = \frac{64\p^2 f_\p^2 (M_\a^2 \sm M_\b^2)}
{3 (M_\a^2 \sm \m^2) (M_\b^2 \sm \m^2)}
\lc 1 \! \pm \! \sqrt{ 1\sp \frac{3 (M_\a^2 \sm \m^2) (M_\b^2 \sm \m^2) \Re \lb L(M_\a^2) \sm L(M_\b^2) \rb}
{32\p^2 f_\p^2 (M_\a^2 \sm M_\b^2)}}\rc \;,
\label{6.13}\\[2mm]
&& c_{\a\a} = - \Re \lb L(M_\a^2) \cos^2 \theta \sp L(M_\b^2) \sin^2\theta \rb + (c_{\a\a} \sm c_{\b\b}) \sin^2 \theta \;,
\label{6.14}\\[2mm]
&& c_{\b\b} = - \Re \lb L(M_\a^2) \cos^2 \theta \sp L(M_\b^2) \sin^2\theta \rb - (c_{\a\a} \sm c_{\b\b}) \cos^2 \theta \;,
\label{6.15}
\eea
\ni
which allow the constants $c_{\a\a}$ and $c_{\b\b}$ to be fixed.

The dependence of the unitarized amplitude $|\bar{A}(s)|^2$ on the mixing angle $\theta$ is shown fig.13,
for the choices $M_\a = 4 \m$ and $M_\b = 8 \m$.

\vspace{3mm}

\ni
\begin{figure}[h]
    \center
    \includegraphics[scale=0.6]{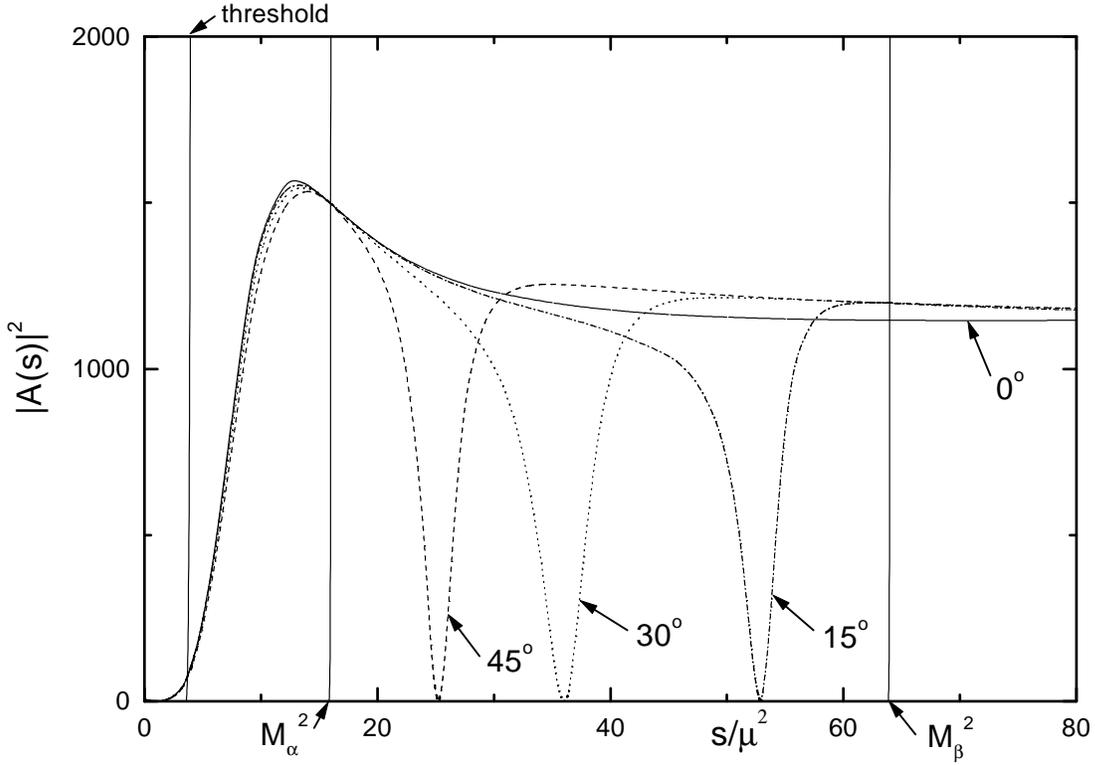}
    \caption{The function $\bar{A}(s)$ is the unitarized amplitude given by
eq.(\ref{6.10}) and the angles quoted represent possible mixings
between resonances $\a$ and $\b$.}
\end{figure}

\vspace{3mm}

\section{summary and general results}

In this work we have used the linear $\s$-model in order to study how chiral symmetry affects
amplitudes incorporating scalar resonances.
Most of our qualitative results confirm, in a lagrangian framework, those derived by  T\"ornqvist\cite{Tor} about ten years ago,
using a unitarized quark model.

One of the implications of chiral symmetry is that the elastic $\p\p$ amplitude must vanish at the subthreshold point $s=\m^2$,
where $\m$ is the pion mass.
As this point is close to the threshold at $s=4\m^2$, the physical amplitude becomes strongly constrained in
the low-energy region.
This aspect of the problem is clearly visible in figs. 6 and 7, for the single resonance case.
From a technical point of view, this happens because the chiral constraint is already present in the tree amplitude,
given by eqs.(\ref{2.6}), (\ref{2.9}) and (\ref{2.10}).
As the unitarization procedure cannot change the position of the chiral zero, it becomes an essential feature of the
full result.

The discussion following eq.(\ref{3.9}) shows that pion loops do affect both the real and imaginary parts of the
denominator of the unitarized amplitude.
However, the neglect of these effects in the real part, which correspond to more complicated expressions, yields
a decent approximation for the full result, as one learns from fig. 8.
Thus, in exploratory studies, one may keep just the pion loop contributions to the imaginary term, which are rather simple.
The single resonance width is given in eq.(\ref{3.13}) and it is worth noting that it incorporates a factor $1/2!$ due
to the exchange symmetry of the intermediate two-pion state.

In section 5 we have produced an extension of the linear $\s$-model aimed at including a second resonance and found out
that the tree $\p\p$ amplitude can be written as

\beq
A_t(s) = \frac{s \sm \m^2}{f_\p^2} \; \lb 1 -\cos^2\theta \,\frac{s \sm \m^2}{s \sm M_\a^2}
-\sin^2\theta \,\frac{s \sm \m^2}{s \sm M_\b^2} \rb \;,
\label{7.1}
\eeq

\ni where $\theta$ is a mixing angle. For $M_\a = M_\b$, one
recovers eq.(\ref{2.6}), for the single resonance case. This
structure gives rise to a second zero for the tree amplitude,
which occurs at a point $s_1$, such that $M_\a^2 < s_1 < M_\b^2$.
The behavior of this zero as a function of $\theta$ can be found
in both eq.(\ref{6.11}) and fig. 13.

When the effects of pion loops over the real part of the amplitude
denominator are neglected, the relationship between the tree and
unitarized amplitudes, given by eq.(\ref{6.10}), becomes
particularly simple:

\beq
\bar{A}(s) = \frac{A_t(s)}{1 - i [3\sqrt{s \sm 4\m^2}\;A_t(s) / 32\p\sqrt{s}]}\;.
\label{7.2}
\eeq

This result, derived in the two-resonance case, is very general and holds for any number of resonances.
It corresponds to the iteration of the tree amplitude as a whole and is not sensitive to its internal structure.
Here, again, the iteration includes a $1/2!$ statistical factor.

In order to extend our results to the case of $N$ coupled scalar resonances, we propose to generalize the
chiral tree amplitude by means of the expression

\beq
A_t(s) = \frac{s \sm \m^2}{f_\p^2} \; \lb 1 - \lambda_1\,\frac{s \sm \m^2}{s \sm M_1^2} - \cdots
-\lambda_N \,\frac{s \sm \m^2}{s \sm M_N^2} \rb \;,
\label{7.3}
\eeq

\ni where the $\lambda_i$ are weights constrained by the condition
$1= \lambda_1 + \cdots \lambda_N$. This amplitude has $N$ zeroes.
The first of them occurs at $s=\m^2$ and is due to chiral
symmetry. The remaining ones are T\"ornqvist zeroes and occur at
the points $s_1, \cdots, s_{N-1}$, between the various resonances.
In principle, the location of these points could be determined
empirically and used to express all the weights $\lambda_i$ as
functions of the masses $M_i$, as in eq.(\ref{6.12}). Feeding this
information back into eqs.(\ref{7.3}) and (\ref{7.1}), one ends up
with an expression for the unitarized amplitude which depends only
on unknown masses, which can be extracted from fits to data.

The results presented in this work were derived in the framework of the linear $\s$-model and,
to some extent, depend on this choice.
On the other hand, they also convey a more general content, namely that the parametrization of the widths of
scalar resonances coupled to pions, associated with the imaginary term in eq.(\ref{7.2}),
must always include a factor $(s\sm \m^2)/f_\p^2$, in order to be compatible with chiral symmetry.

At present, we are considering the inclusion of $K$ and $\eta$ mesons in our results.

\clearpage
\ni
{\bf Acknowledgement}\\
It is our pleasure to thank Ignacio Bediaga for several
conversations about experimental aspects of scalar resonances.


\end{document}